\documentclass{aa}
\usepackage{graphics}

\usepackage{epsfig}
\usepackage{latexsym}
\usepackage{natbib}
\def\E$\gamma${E_\$\gamma$}

\def\deg       {$^{\circ}$}
\def \ga       {$\gamma$}
\def \sig      {$\sigma$}
\def \gray     {$\gamma$-ray}
\def \grays    {$\gamma$-rays}
\def \phibar   {$ \bar\varphi$}

\begin{document}

%\hyphenation{brems-strah-lung}

%\thesaurus{   
%               03                      % Main Journal (?) 
%              (13.07.2;                % Gamma rays: observations
%               11.01.2;                % Galaxies: active
%               11.17.4 PKS1622-297);   % Galaxies: quasars: individual: 1622
%              }
%

\title{COMPTEL Observations of the Gamma-Ray Blazar PKS~1622-297}

\author{ S.~Zhang\inst{1,6},
 	 W.~Collmar\inst{1},
         K.~Bennett\inst{4},
         H.~Bloemen\inst{2},
         W.~Hermsen\inst{2},
         M.~McConnell\inst{3}, 
         O.~Reimer\inst{5},
         V.~Sch\"onfelder\inst{1},
         S. J. Wagner\inst{7},
         and O.R.~Williams\inst{4}
}
 
\offprints{S.~Zhang}
\institute{Max-Planck-Institut f\"ur extraterrestrische Physik,
               P.O. Box 1603, D-85740 Garching, Germany 
            \and
             SRON National Institute for Space Research, Sorbonnelaan 2, NL-3584 CA Utrecht,
               The Netherlands
            \and
               University of New Hampshire, Institute for the Study
               of Earth, Oceans and Space, Durham NH 03824, USA
            \and
              Astrophysics Division, Space Science Department of
               ESA/ESTEC, NL-2200 AG Noordwijk, The Netherlands
            \and
              NASA/Goddard Space Flight Center, Greenbelt, MD 20771, USA
	     \and
	      Laboratory of Cosmic Ray and High Energy Astrophysics, 
              Institute of High Energy Physics,
	     P.O.Box 918-3, Beijing 100039, China
            \and
              Landessternwarte Heidelberg, K\"onigstuhl,
              D-69117 Heidelberg, Germany
          }

\mail{shz@mpe.mpg.de}

\date{Received November 6, 2001  / Accepted February 19, 2002}

\titlerunning{COMPTEL Observations ...}
\authorrunning{S.~Zhang et al.}

\abstract{
We report results of observations and analyses on the 
\gray\ blazar PKS~1622-297, with emphasis on
the COMPTEL data (0.75 - 30~MeV) collected between April 1991 and 
November 1997.
PKS 1622-297 was detected as a source of \grays\ by the EGRET
experiment aboard CGRO in 1995 during a \gray\ outburst
at energies above 100 MeV lasting for five weeks.
 In this time period the blazar was significantly 
($\sim$ 5.9 \sig) detected by COMPTEL at 10-30 MeV.
At lower COMPTEL energies the detection is marginal,
resulting in a hard MeV spectrum.
 The combined COMPTEL/EGRET energy spectrum shows a break 
at MeV energies. The 
broad-band spectrum (radio - \grays) shows that the \gray\
emission dominates the overall power output. 
On top of the 5-week \gray\ outburst, EGRET detected a huge flare 
lasting for $>$ 1 day. Enhanced MeV emission (10 - 30~MeV) is found
near the time of this flare, suggesting a possible time delay with
respect to the emission above 100~MeV. Outside the 5-week flaring period
in 1995, we do not detect MeV emission from PKS~1622-297.

\keywords{\ga\ rays: observations --- galaxies: active --- galaxies: quasars: individual: PKS~1622-297}

}

\maketitle

\section{Introduction}

The EGRET experiment aboard the Compton Gamma-Ray Observatory (CGRO)
has detected roughly 90 blazar-type Active Galactic Nuclei (AGN)
at \gray\ energies above 100~MeV (Hartman et al., 1999).  
Ten of these were also detected by COMPTEL measuring at lower-energy
\grays\ in the 0.75 - 30~MeV band, e.g.
CTA~102 and 3C~454.3 (Blom et al., 1995a), PKS~0528+134 
(Collmar et al., 1997a), PKS~0208-512 (Blom et al., 1995b), 3C~273 and 3C~279 (Hermsen et al., 1993; Williams et al., 1995; Collmar et al., 2000).
They are often detected during \gray\ flaring periods reported by 
EGRET at energies above 100~MeV.
Another such example is the blazar PKS~1622-297 discussed in this paper. 
It was detected by 
the CGRO experiments OSSE, COMPTEL and EGRET at \gray\ energies during
a 5-week flaring period between
June 6 and July 10, 1995 (Kurfess et al., 1995; Collmar et al., 1997b;
Mattox et al., 1997).

The quasar PKS~1622-297 has a redshift of z = 0.815 (Wright \& Otrupcek, 1990)
and is located  at 
(l,b) = (348.82\deg,13.32\deg).
It is a known radio source with a flat radio spectrum (flat-spectrum 
radio quasar) (K\"uhr et al., 1981; Steppe et al., 1993), showing polarization of
4.6$\%$ at 5 GHz (Impey \& Tapia, 1990). 
In the optical band PKS~1622-297 is a weak source with typical visual
magnitudes around 20 to 21~mag (e.g. Torres \& Wroblewski, 1984; 
Saikia et al., 1987). X-rays of PKS 1622-297 were detected by ROSAT with a flux of (3.2$\pm$0.8)$\times$10$^{-13}$ erg~cm$^{-1}$~s$^{-1}$ in the 0.1-2.4 keV energy band (Mattox et al., 1997).

In June and July 1995 the quasar was observed to be flaring
at \gray\ energies above 100 MeV by EGRET.
The COMPTEL experiment also detected the source 
at MeV \gray\ energies in this period. Preliminary COMPTEL results,
reported by Collmar et al. (1997b), provide a source detection
at a significance level of $\sim$ 5 \sig\ and a hint
for flux variability in the energy range 10-30 MeV.
After the \gray\ outburst was recognized a target-of-opportunity 
(ToO) observation for CGRO was declared (June 30 - July 10,
1995) which caused the OSSE experiment 
(0.05 - $\sim$10~MeV) to participate in observing PKS~1622-297.
The quasar was detected by OSSE at $\sim$~6~\sig.
The simultaneous EGRET and OSSE spectra are consistent with a simple power-law shape with photon index 1.87
(Mattox et al., 1997).
On top of this \gray\ outburst, EGRET observed a huge \gray\ flare
with intraday variability,
which occurred around June 24, 1995 and lasted for $>$ 1 day
(Mattox et al., 1997).
Its peak flux of (17$\pm$3)$\times$10$^{-6}$~ph~cm$^{-2}$~s$^{-1}$ 
is among 
the brightest ones ever observed from any blazar by EGRET.  
%%%%% Table 1 %%%%%%%%%%%
\begin{table*}[thb]
\caption{COMPTEL observations of PKS 1622-297 during the first 6 years of the 
CGRO mission. The CGRO VPs, their time periods, prime observational targets, offset angles, effective exposure and the CGRO Phases  are given.}
\begin{flushleft}
\begin{tabular}{cccccc}
\hline 
\multicolumn{1}{c}{VP}&\multicolumn{1}{c}{Date}&\multicolumn{1}{c}{Object}&\multicolumn{1}{c}{Offset angle}&\multicolumn{1}{c}{Effective exposure }&\multicolumn{1}{c}{CGRO Phases}\\ 
\multicolumn{1}{c}{}&\multicolumn{1}{c}{(dd/mm/yy)}&\multicolumn{1}{c}{}&\multicolumn{1}{c}{}&\multicolumn{1}{c}{days}&\multicolumn{1}{c}{}\\ \hline 
5.0 & 12/07/91-26/07/91& Gal. Center & 20.6\deg& 5.0&Phase I\\  
16.0 & 12/12/91-27/12/91 &  SCO X-1 & 12.8\deg& 5.6&\\ 
23.0 & 19/03/92-02/04/92 & CIR X-1 & 28.3\deg& 2.2&\\ 
27.0 & 28/04/92-07/05/92 & 4U 1543 & 19.7\deg& 1.7&\\ \hline
210.0 & 22/02/93-25/02/93& Gal. Center & 9.7\deg& 0.6&Phase II\\  
214.0 & 29/03/93-01/04/93 &  Gal. Center & 9.7\deg& 0.7&\\  
219.4 & 05/05/93-06/05/93 & Gal. Center & 2.68\deg& 0.5&\\ 
223.0 & 31/05/93-03/06/93 & Gal. Center & 16.9\deg& 0.7&\\ 
226.0 & 19/06/93-29/06/93& GAL 355+05 & 10.3\deg& 2.1&\\  
229.0 & 10/08/93-11/08/93 &  GAL 5+05 & 18.0\deg& 0.2&\\  
232.0 & 24/08/93-26/08/93 & GAl 348+00 & 13.4\deg& 0.4&\\ 
232.5 & 26/08/93-07/09/93 & GAl 348+00 & 13.4\deg& 2.6&\\ \hline 
302.3 & 09/09/93-21/09/93& GX 1+4 & 13.0\deg& 2.6&Phase III\\  
323.0 & 22/03/94-05/04/94 &  GAL 357-11 & 25.9\deg& 4.3&\\  
324.0 & 19/04/94-26/04/94 & GAl 015+05 & 26.9\deg& 2.5&\\ 
334.0 & 18/07/94-25/07/94 & GAL 009-08 & 29.5\deg& 2.1&\\ 
336.5 & 04/08/94-09/08/94& GRO J1655-40 & 13.4\deg& 1.5&\\  
338.0 & 29/08/94-31/08/94 &  GRO J1655 & 11.5\deg& 0.6&\\  \hline
414.3 & 29/03/95-04/04/95 & GRO J1655-40 & 12.8\deg& 2.0&Phase IV\\ 
421.0 & 06/06/95-13/06/95& Gal. Center & 14.5\deg& 2.2&\\ 
422.0 & 13/06/95-20/06/95 &  Gal. Center & 15.2\deg& 2.1&\\ 
423.0 & 20/06/95-30/06/95 & Gal. Center & 19.2\deg& 3.2&\\ 
423.5 & 30/06/95-10/07/95 & PKS 1622-297 & 3.0\deg& 2.8&\\ \hline
508.0 & 14/12/95-20/12/95 & GAL 005+00 & 21.3\deg& 1.7&Phase V\\ 
516.1 & 18/03/96-21/03/96& GRO J1655-40 & 10.9\deg& 0.8&\\  
524.0 & 09/07/96-23/07/96 &  GX 339-4 & 17.8\deg& 3.8&\\  
529.5 & 27/08/96-06/09/96 & GRO J1655-40 & 11.5\deg& 2.7&\\ \hline
624.1 & 04/02/97-11/02/97 & Gal 16+00 & 28.5\deg& 2.0&Phase VI\\ 
\hline
\end{tabular}\end{flushleft}
\label{tab1}
\end{table*}
%%%%% Table 1 %%%%%%%%%%%

During the CGRO mission (April 1991 to June 2000), COMPTEL 
has conducted many observations of PKS 1622-297. 
We used the first 6 years of data to study the MeV behaviour of PKS~1622-297  
(i.e., the data before the second reboost of CGRO in 1997
which changed the background environment for COMPTEL).
We describe the COMPTEL observations in Sect.~3. 
In Sect.~4 we report the main COMPTEL results on detections,
light curves and energy spectra, and compare them to the results 
in neighboring energy bands, in particular to the EGRET ones.
We discuss our findings in the framework of current \gray\
emission scenarios for blazars in Sect. 5, and finally 
summarize in Sect. 6.

%######################################################################
\section{Instrument and data analysis}
%######################################################################
The imaging Compton Telescope COMPTEL is sensitive to 
\grays\ in 
the energy range 0.75-30 MeV with an energy-dependent energy resolution
between 5$\%$ and 10$\%$ and an angular resolution of 1\deg\ - 2\deg.
It consists of two layers of detectors (upper and lower)
 which are separated by 1.58~m. An incident \ga\ photon is 
first Compton scattered in one of the upper detectors and then 
-- ideally -- completely absorbed in one of the lower detectors. 
Two orthogonal coordinates ($\chi$, $\psi$), describing the direction
of the scattered \gray, can be obtained from the 
interaction locations in the two detector layers.
The Compton scatter angle, \phibar,  
 can be calculated from the 
measured energy deposits in the two detectors.
Therefore the direction of the incident photon is located
on a projected circle on the sky.
The scatter direction ($\chi$, $\psi$) and the 
Compton scatter angle (\phibar) constitute a three-dimensional
data space in which the spatial response of the instrument
is cone-shaped. For more 
details about the COMPTEL instrument see Sch\"onfelder et al. (1993).

There are two standard imaging methods, maximum-likelihood and  maximum entropy, for the COMPTEL data analysis in this three-dimensional data 
space. Flux estimates and statistical significances 
can be obtained by applying the maximum-likelihood method. The 
quantity -2ln$\lambda$, where $\lambda$ is the
 likelihood ratio of two hypotheses
(background, background + source) provides the detection significance 
for a \gray\ source (de Boer et al., 1992). 
This -2ln$\lambda$ quantity has a $\chi_{3}^{2}$ distribution
(3 degrees of freedom) for a unknown point source 
(search modus) and a $\chi_{1}^{2}$ distribution for a known source
 (check at a
given position). In order to estimate the COMPTEL background the standard filter
technique (Bloemen et al., 1994) was applied. 
Because PKS~1622-297 is located towards the inner Galaxy, models 
of the diffuse galactic emission 
(bremsstrahlung, inverse-Compton radiation) were included in the analysis.
An isotropic component describing the extragalactic
diffuse emission was included as well.
For flux estimates we assumed point spread functions with an E$^{-2}$
power law shape for the source input spectrum.
We note that the derived fluxes are weakly dependent on this particular
choice.

%######################################################################
\section{Observations}
%######################################################################
From the beginning of the CGRO mission in April 1991 up to the 
second reboost of CGRO in April 1997, which changed the
background environment 
at lower ($<$4.3~MeV) COMPTEL energies significantly,
PKS 1622-297 was many times within the COMPTEL field-of-view (fov).
This time period covered 230 CGRO viewing periods (VP) typically
lasting 1 to 2 weeks each, and 6 so-called CGRO Phases each covering 
a time period of roughly 1 year.   
For our analysis we selected 28 pointings (see Table~1) for which 
the angle between the COMPTEL pointing direction and the source direction
was less than 30\deg. As listed in Table~1, CGRO viewing period (VP) 229.5
is added to VP 229.0 because the pointing was the same.
In summary, between April 1991 and February 1997 PKS~1622-297
was 215 days within 30\deg\ of the COMPTEL pointing direction, resulting in
 a total effective exposure (100\% COMPTEL pointed directly to the source)
of 59.2 days.

%%%%% Table 2 %%%%%%%%%%%
\begin{table}[bh]
\caption{Fluxes and flux limits of PKS 1622-297 for the individual 
'flaring' VPs, the sum of '\gray\ flaring' (VPs 421 to 423.5)
and '\gray\ quiescent' states. The flux units are 
10$^{-5}$~ph~cm$^{-2}$~s$^{-1}$. The energy bands are given in MeV.
 The errors are 
1 \sig\ and the upper limits are 2 \sig.} 
\begin{flushleft}
\begin{tabular}{ccccc}
\hline\noalign{\smallskip}
  Period & 0.75-1 & 1-3 & 3-10 & 10-30   \\
\hline\noalign{\smallskip}
VP 421.0 & $<$23.5 & $<$22.3 & $<$10.3 & 3.2$\pm$1.5 \\
VP 422.0 & $<$50.6 & $<$20.1 & $<$7.8 & 2.7$\pm$1.4      \\
VP 423.0 & $<$38.4 & $<$23.6 & $<$9.4 & 5.9$\pm$1.4  \\
VP 423.5 & $<$18.7 & $<$16.3 & 4.2$\pm$3.0 & 1.8$\pm$1.0 \\
Flaring state & $<$10.1 & $<$9.7 & 2.5$\pm$1.7 & 3.5$\pm$0.6 \\
Quiescent state &$<$6.0 &$<$4.4 & $<$1.7& $<$0.8\\
\hline\noalign{\smallskip}
\end{tabular}\end{flushleft}
\label{tab2}
\end{table}
%%%%% Table 2 %%%%%%%%%%%

\section{Results}
%######################################################################
\subsection{Source detections}

EGRET reported a strong detection of 25\sig\  
with a large \gray\ flux during the \gray\ outburst period in 1995
(Mattox et al., 1997). Outside this period only marginal evidence for the 
source, i.e. a low flux level, is found by EGRET (Fig.~1).
According to the measured EGRET flux, 
we define two \gray\ states for PKS 1622-297:
a '\gray\ flaring' state covering the 4 VPs 421.0 to 423.5,
and a '\gray\ quiescent' state which covers the remaining 24 pointings.
In the rest of the paper, we will refer to the relevant observational
periods/times as the '\gray\ flaring' and '\gray\ quiescent' 
periods. 

%%%%% Fig 1 %%%%%%%%%%%%%%%%%%%%%%%%%%%%%%%%%%%%%%%%%%%%
\begin{figure}[tb]
\centering
\includegraphics[width=8.0cm]{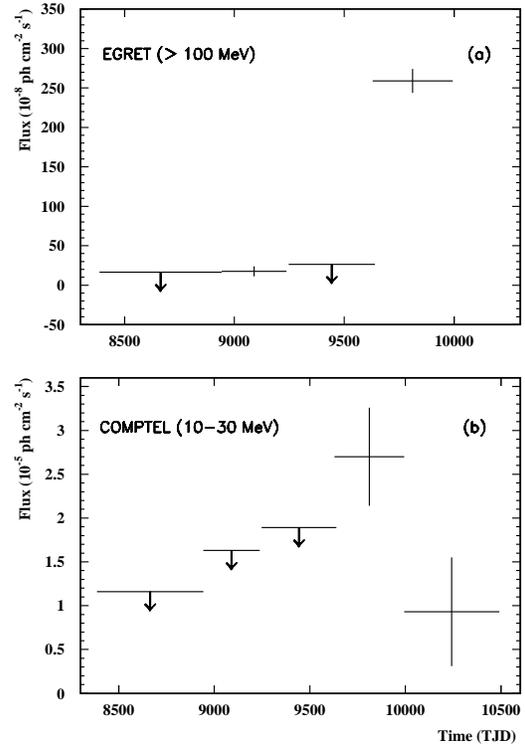}
\caption{\gray\ light curves of PKS 1622-297 as observed by EGRET
 at energies $>$ 100 MeV (a) and by COMPTEL in the 
energy range 10-30 MeV 
(b) for individual Phases 
(Phases 1-4 for EGRET and Phases 1-6 for COMPTEL). 
Phase 5 and Phase 6 are combined into one bin because
 only one COMPTEL observation
 (VP 624.1) from Phase 6 before the second reboost was
 selected in our analysis.   
 The EGRET data are from Hartman et al. (1999).
 The error bars are 1 \sig\ and the upper limits are 2 \sig.
\label{fig:lightcurve0}}
\end{figure}
%%%%% Fig 1 %%%%%%%%%%%%%%%%%%%%%%%%%%%%%%%%%%%%%%%%%%%%

\begin{figure}[tb]
\centering
\includegraphics[width=9.0cm]{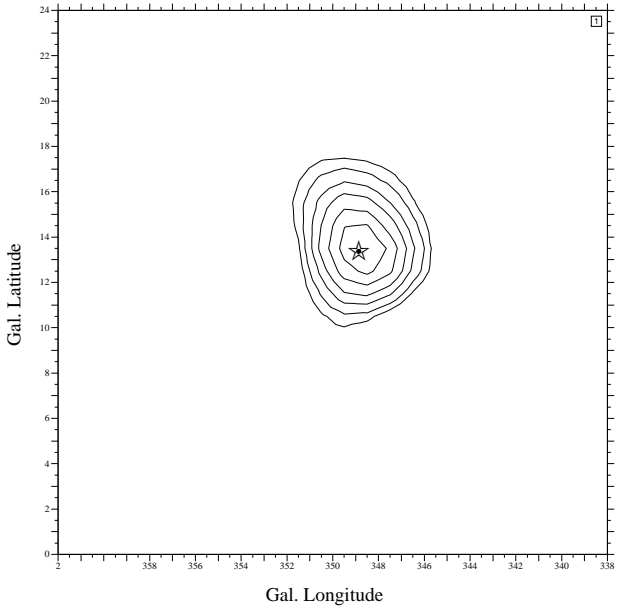}
\caption{COMPTEL 10-30 MeV map of the 5-week \gray\ 
high state of PKS 1622-297 ($\star$) 
in 1995. The contour lines start at a detection significance level
 of 3 \sig\ with steps of 0.5 \sig.
\label{fig:skymap}}
\end{figure}

COMPTEL imaging analysis revealed evidence for PKS~1622-297
during the 5-week flaring period in 1995. 
In the sum of these data the quasar is 
detected in the uppermost COMPTEL band (10 - 30~MeV) with 
a significance of 5.9 \sig, assuming $\chi_{1}^{2}$-statistics
for a known source (Fig.~\ref{fig:skymap}). 
Below 10~MeV the detection of the source is only marginal ($\sim$~1.5~\sig\
in the 3-10~MeV band) 
or non-detections occur (below 3~MeV). Evidence for the source 
in the 10-30~MeV band is found in each of the 4 individual VPs, 
typically at the a 2\sig-significance level.
Outside this \gray\ flaring period in 1995 PKS~1622-297 is detected marginally by COMPTEL only in Phase 5+6 in the 10-30 MeV band (Fig.~1). 
The relevant flux results are given in Table~2.

%%%%% Fig 2 %%%%%%%%%%%%%%%%%%%%%%%%%%%%%%%%%%%%%%%%%%%%
%%%%% Fig 2 %%%%%%%%%%%%%%%%%%%%%%%%%%%%%%%%%%%%%%%%%%%%

\subsection{Time variability}
\begin{figure}[tb]
\centering
\includegraphics[width=8.0cm]{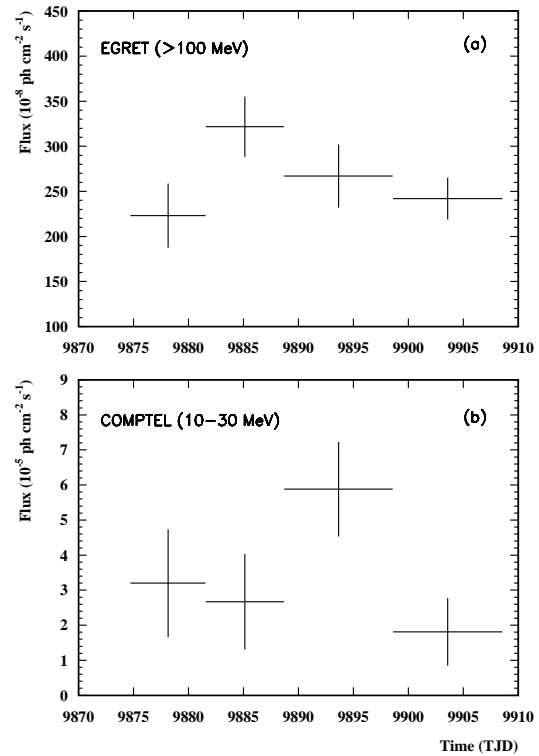}
\caption{\gray\ light curves of PKS 1622-297 as observed by EGRET
 at energies $>$ 100 MeV (a) and by COMPTEL in the 
energy range 10-30 MeV (b) for the 4 individual
 VPs of the 5-week \gray\ high state in 1995. 
 The EGRET data are from Hartman et al. (1999).
 The error bars are 1 \sig.
 \label{fig:lightcurve1}}
\end{figure}

During the first 6 years of the CGRO mission, 
PKS~1622-297 was strongly variable at \gray\ energies above 100 MeV
with the major outburst in 1995.
Due to the poorer statistics, the detection of time variability cannot 
easily be claimed by COMPTEL.
Nevertheless, the COMPTEL detections and non-detections in the
10-30 MeV band still follow the EGRET trend on long  (Fig.~1)
as well as on short time scales (Fig.~\ref{fig:lightcurve1}).
On long terms (years) the maximum flux is reached during CGRO 
Phase 4 (October 1994 to October 1995), which comprises the \gray\
outburst period in 1995.
On shorter time scales (weeks), during the \gray\ flare state, 
evidence for PKS~1622-297 is found in each VP by EGRET as well
as by COMPTEL. 
Both flare-state light curves do not show strong flux variability  
(Fig.~\ref{fig:lightcurve1}).

%%%%% Fig 3 %%%%%%%%%%%%%%%%%%%%%%%%%%%%%%%%%%%%%%%%%%%%
%%%%% Fig 3 %%%%%%%%%%%%%%%%%%%%%%%%%%%%%%%%%%%%%%%%%%%%

%%%%% Fig 4 %%%%%%%%%%%%%%%%%%%%%%%%%%%%%%%%%%%%%%%%%%%%
\begin{figure}[tb]
\centering
\includegraphics[width=8.0cm]{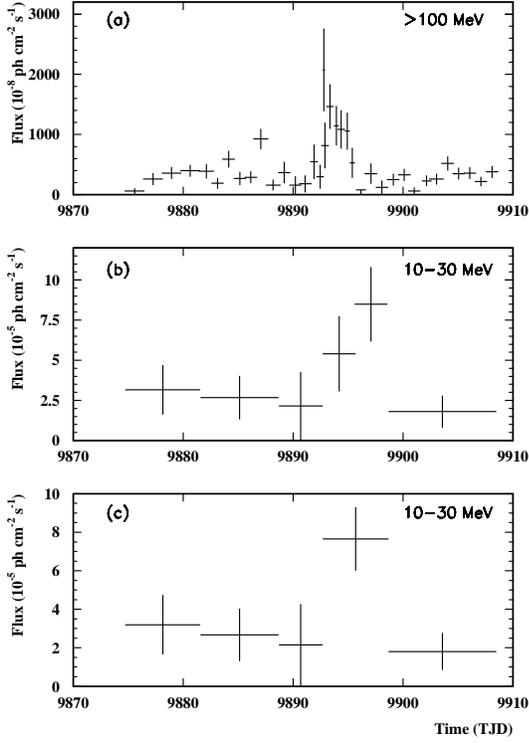}
\caption{High time-resolution EGRET light curve of PKS 1622-297 (a)
compared to the COMPTEL light curves (b, c), where the major
 flaring period is better 
resolved in time.  Fig. 3c is the same as Fig. 3b 
with the two high fluxes combined to show the enhanced 
MeV emission around the time 
of the major EGRET flare. The EGRET data are from Mattox et al. (1997).
 The error bars are 1 \sig.
\label{fig:lightcurve2}}
\end{figure}
%%%%% Fig 4 %%%%%%%%%%%%%%%%%%%%%%%%%%%%%%%%%%%%%%%%%%%%

On even shorter time scales (days), however, Mattox et al. (1997) 
report a short huge flare on top 
of the \gray\ high state light curve, which occurred 
within VP 423.0 (flux maximizing at TJD 9892.77) and
 lasted for $>$ 1 day (Fig.~\ref{fig:lightcurve2}). 
We searched the COMPTEL data for a possible MeV counterpart 
by subdividing VP 423 into 3 parts according
to the EGRET light curve: a pre-flare (TJD 9888.8 - 9892.7), on-flare
 (TJD 9892.7 - 9895.6), and post-flare part (TJD 9895.6 - 9898.6).
The on-flare part covers the time period of TJD 9892.7-9893.7, during which 
the short huge flare peaked in flux reaching a value of
$\sim$ (17$\pm$3)$\times$10$^{-6}$~ph~cm$^{-2}$~s$^{-1}$ 
at energies $>$ 100 MeV (Mattox et al., 1997).
The 10-30 MeV light curve is shown in Fig.~\ref{fig:lightcurve2}.
It is obvious that the MeV emission is stronger in the flare 
and post-flare period than during other parts. 
The combination of on-flare and post-flare periods provides a
$\sim$ 5 \sig\ detection of the source and clearly shows enhanced
 MeV emission around the time of the major EGRET flare (Fig.~\ref{fig:lightcurve2}).
Fitting three data points, pre-flare, combination of on-flare
and post-flare, and the flux for VP 423.5 by assuming a constant flux, 
results in a $\chi_{min}^{2}$-value of 9.4, which, with 2 
degrees of freedom, corresponds to a probability of 
$\sim$ 9$\times$10$^{-3}$ for a constant flux or roughly 2.6 \sig\ for 
a variable flux. 
We note that the
highest COMPTEL flux is found immediately after the flare in the EGRET band
(Fig.~\ref{fig:lightcurve2}). This suggests a
time delay of a few days between the
two bands in the sense that the higher energies \grays\ come first. 

\subsection{Energy spectra}
\begin{figure}[tb]
\centering
\includegraphics[width=8.0cm]{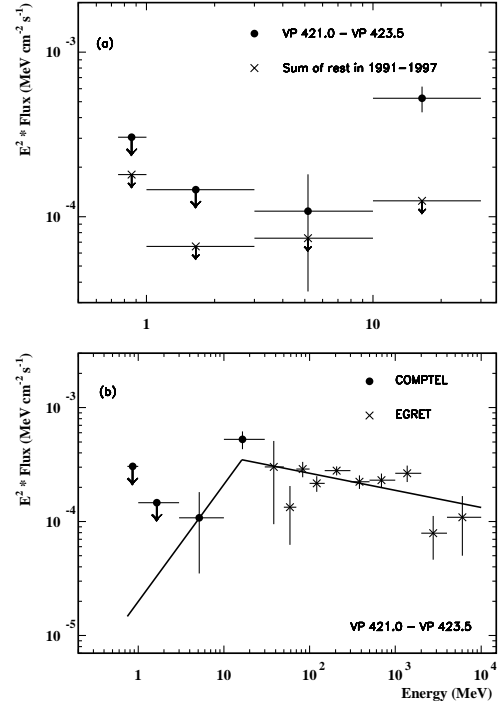}
\caption{The COMPTEL spectra of the 5-week flaring state and
 the sum of the rest between 1991 and 1997 (a) as well as 
the combined COMPTEL/EGRET spectrum for the flaring state (b).
 The error bars are 1 \sig\ and the upper limits are 2 \sig.
\label{fig:quie_flare_spec}}
\end{figure}

We generated two separate MeV spectra of PKS~1622-297: a '\gray\ flaring'
spectrum by combining the COMPTEL data of the 4 relevant VPs
and a '\gray\ quiescent' spectrum by combining the rest of the
observations (24 VPs). 
A fit to the flaring spectrum results in a photon index $\alpha$ of
0.65$^{+0.53}_{-2.49}$ (Table~3). Due to the marginal detections
at energies below 10~MeV the spectral index cannot be constrained well. 
However, a 'hard' ($\alpha_{ph}$$<$2, $\sim$$E^{-\alpha}$) MeV shape
is obvious (Fig.~\ref{fig:quie_flare_spec}). 
This shows that during the \gray\ outburst period the 
MeV band is still on the rising part of the IC emission. 
No meaningful conclusions can be derived for the spectrum of the 
quiescent state, because only upper limits 
were obtained. A comparison of both spectra shows that during 
the flaring state the 10-30~MeV flux was at least 4 times larger
than during the rest. 
 
 %%%%% Table 3 %%%%%%%%%%%
\begin{table*}[thb]
\caption[]{Results of the power-law fit (the upper panel)
 and broken power-law fit (the lower panel) for the 
flaring state of PKS~1622-297. 
The errors are derived by the $\chi^{2}_{min}$ + 2.3 contour
level for fitting two parameters of power law or broken power law
(with $\alpha_{2}$ and I$_{0}$ fixed), 
and by the $\chi^{2}_{min}$ + 4.7 contour
level for fitting four parameters of broken power law. }
\begin{flushleft}
\begin{tabular}{ccccc}
\hline\noalign{\smallskip}
Data & PL-Index & I$_0$ & E$_0$ & $\chi^{2}_{red}$  \\
$\#$& ($\alpha$) & (10$^{-8}$cm$^{-2}$ s$^{-1}$ MeV$^{-1}$) & (MeV) &   \\ 
\hline\noalign{\smallskip}
VPs 421.0-423.5 (0.75 MeV - 30 MeV) & 0.65$^{+0.53}_{-2.49}$ & 588.0$^{+914.0}_{-451.0}$  & 3.0 & 0.34  \\  
VPs 421.0-423.5 (30 MeV - 10 GeV) & 2.15$^{+0.07}_{-0.07}$ & 0.26$^{+0.02}_{-0.02}$ & 294.8 & 2.56  \\   
VPs 421.0-423.5 (0.75 MeV - 10 GeV) & 2.03$^{+0.04}_{-0.04}$ & 0.54$^{+0.05}_{-0.05}$ & 200.0 & 4.26  \\   
VP 423.5 (0.05 MeV - 10 GeV) & 1.84$^{+0.02}_{-0.02}$  & 1.6$^{+0.2}_{-0.2}$ & 100.0 &  1.31  \\   
\hline\noalign{\smallskip}
\end{tabular}\end{flushleft}
\begin{flushleft}
\begin{tabular}{cccccc}
\hline\noalign{\smallskip}
Data & $\alpha_{2}$ & I$_0$ (200 MeV) & $\delta\alpha$ & E$\rm_b$ & $\chi^{2}_{red}$  \\
$\#$&  &  & & (MeV) &   \\ 
\hline\noalign{\smallskip}
VPs 421.0-423.5 (0.75 MeV - 10 GeV) & 2.15   & 0.60  & 1.18 & 16.2$^{+3.3}_{-9.9}$& 2.06  \\ 
 & (fixed)  & (fixed)  &  & &   \\ \hline   
VP 423.5 (0.05 MeV - 10 GeV) & 2.57$^{+1.04}_{-0.60}$  & 1.49$^{+7.30}_{-1.00}$ & 0.78$^{+0.47}_{-0.59}$ & 630.0$^{+1504.0}_{-622.0}$& 0.67  \\   
\hline\noalign{\smallskip}
\end{tabular}\end{flushleft}

\label{tab3}
\end{table*}
%%%%% Table 3 %%%%%%%%%%%
  
The EGRET spectrum (30~MeV to 10~GeV) of the \gray\ flaring state 
is consistent with a power-law shape and a photon 
index of 2.15$\pm$0.07 (Table~3, Fig.~5). If combined with the
simultaneous COMPTEL spectrum a spectral 
break becomes evident.

Fitting a simple power-law model,

\begin{equation}
I(E) = I_{0}  (E/E_{0})^{-\alpha} \,\, {\rm photon\,\,cm}^{-2} {\rm s}^{-1} {\rm MeV} ^{-1}
\end{equation}  

where the parameter $\alpha$ is the photon index, and I$_0$ the 
differential flux 
at the normalization energy E$_0$, results in an unacceptable  
reduced $\chi^{2}$-value of 4.26 (Table~3).
To improve on that we applied a broken power-law model of the form

\begin{equation}
I(E) = \left\{ \begin{array}{ll}
 I_{0}  (E/E_{0})^{-\alpha_{2}} &\mbox {if $E>E_{b}$} \\
 I_{0}  (E_{b}/E_{0})^{-\alpha_{2}} \, (E/E_{b})^{(\Delta\alpha - \alpha_{2})}
                                 &\mbox {if $E<E_{b}$}
\end{array}
\right.
\end{equation}  

where  $\alpha_{2}$ is the high-energy spectral index,
$\Delta\alpha$ the break in the spectral photon index towards
lower energies ($\Delta\alpha$ = $\alpha_{2}$ - $\alpha_{1}$),
and E$\rm_b$ the break energy. 
With a reduced $\chi^{2}$ of 2.1 this model significantly improves the 
fit, proving quantitatively the spectral turnover.
By fixing the parameters I$_0$ and $\alpha_{2}$ to the values
obtained by fitting a power-law model to the EGRET data only,
we derive information on the spectral 
break. Best-fit values of $\Delta\alpha$ = 1.18
and E$\rm_b$ = 16.2$^{+3.3}_{-9.9}$ MeV are derived. 
The 1 \sig\ errors are obtained by adding 2.3 to the minimum
$\chi^{2}$-value as is appropriate for 2 parameters of interest 
(Lampton et al., 1976). 
However, due to the insignificance of the COMPTEL data at energies below
10 MeV the low-energy slope ($\alpha_{1}$) could not be constrained.

%%%%% Fig 5 %%%%%%%%%%%%%%%%%%%%%%%%%%%%%%%%%%%%%%%%%%%%
%%%%% Fig 5 %%%%%%%%%%%%%%%%%%%%%%%%%%%%%%%%%%%%%%%%%%%%

%%%%% Fig 6 %%%%%%%%%%%%%%%%%%%%%%%%%%%%%%%%%%%%%%%%%%%%
\begin{figure}[tb]
\centering
\includegraphics[width=8.0cm]{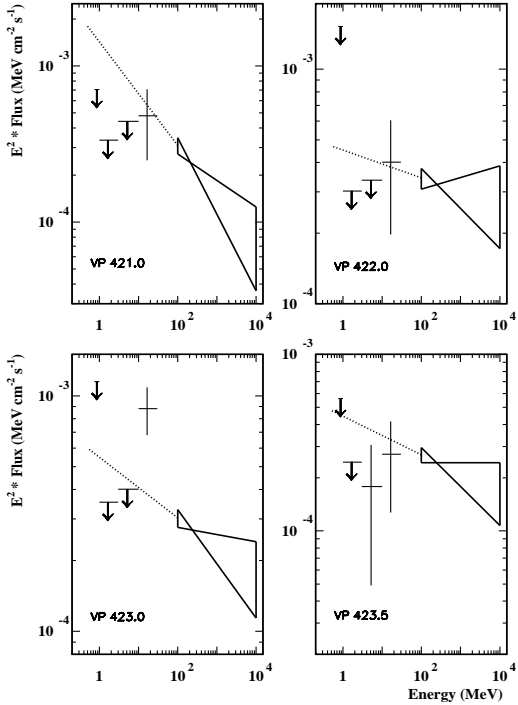}
\caption{The combined COMPTEL/EGRET spectra of the 4 individual
 VPs in the flaring state.
 The data points
represent the COMPTEL fluxes, the dotted lines 
the extrapolations of the EGRET power-law spectra and the solid lines the EGRET spectra showing 1 \sig\ error in their spectral indices.
 The error bars are 1 \sig\ and the upper limits are 2 \sig.
\label{fig:4spectra}}
\end{figure}
%%%%% Fig 6 %%%%%%%%%%%%%%%%%%%%%%%%%%%%%%%%%%%%%%%%%%%%

%%%%% Fig 7 %%%%%%%%%%%%%%%%%%%%%%%%%%%%%%%%%%%%%%%%%%%%
\begin{figure}[tb]
\centering
\includegraphics[width=8.0cm]{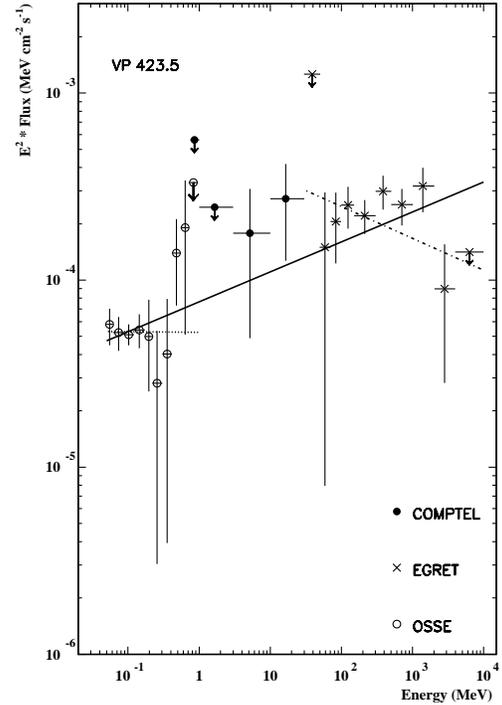}
\caption{The combined simultaneous OSSE/COMPTEL/EGRET spectrum for 
the ToO period (VP 423.5).
The EGRET and OSSE data are from Mattox et al. (1997). The dotted line symbolizes the OSSE fit and the dashed-dotted line 
the EGRET fit (Mattox et al., 1997), the solid line the combined OSSE/COMPTEL/EGRET fit. 
 The error bars are 1 \sig\ and the upper limits are 2 \sig.
\label{fig:vp423.5_spec}}
\end{figure}
%%%%% Fig 7 %%%%%%%%%%%%%%%%%%%%%%%%%%%%%%%%%%%%%%%%%%%%

The EGRET spectra of the 4 flare-state VPs have been analyzed
individually assuming power-law shapes. Within uncertainties the
4 derived photon indices are identical (Table~4).
We then have combined the EGRET and COMPTEL spectra of these 4 
VPs to check for possible spectral variability during the \gray\ 
outburst over a broader energy band (Fig.~\ref{fig:4spectra}).
The COMPTEL 10-30 MeV flux follows for each VP generally the
extrapolation of the EGRET spectrum except from VP~423,
where the flux is slightly above the EGRET extrapolation.
The non  or marginal detections at energies below 10 MeV
indicate a hard MeV spectrum and -- together with
the EGRET spectral shape --
a spectral break at MeV energies for VP 421,
 even taking into account the errors on the spectral index (Fig.~\ref{fig:4spectra}).

 %%%%% Table 4 %%%%%%%%%%%
\begin{table}[thb]
\caption[]{Photon spectral index, obtained from our reanalysis of
the EGRET data, and the integral flux (E $>$ 100 MeV, from Hartman et al., 1999)
 for the 4 individual VPs of the flaring state.}
\begin{flushleft}
\begin{tabular}{ccc}
\hline\noalign{\smallskip}
Observations & Photon Index & Integral flux  \\
Periods& ($\alpha$) & ($\times$ 10$^{-8}$~ph~cm$^{-2}$~s$^{-1}$)   \\ 
\hline\noalign{\smallskip}
VP 421.0 & 2.33$\pm$0.16  & 233.1$\pm$35.7 \\  
VP 422.0 & 2.06$\pm$0.11  & 321.8$\pm$33.5 \\ 
VP 423.0 & 2.13$\pm$0.10  & 267.1$\pm$34.9 \\
VP 423.5 & 2.11$\pm$0.11 & 242.1$\pm$23.2  \\  
\hline\noalign{\smallskip}
\end{tabular}\end{flushleft}
\label{tab4}
\end{table}
%%%%% Table 4 %%%%%%%%%%

After the \gray\ activity of PKS~1622-297 had been recognized
by EGRET, a target-of-opportunity (ToO) observation for CGRO 
was declared and carried out in VP~423.5 with participation of OSSE. 
This provided a measurement of the spectrum down to 0.05~MeV. 
Mattox et al. (1997) reported that the OSSE and
EGRET data can each  be represented by similar power-law shapes with  
photon indices of 2.0$\pm$0.2 and 2.2$\pm$0.1, respectively (Fig.~\ref{fig:vp423.5_spec}).
However, fitting a simple power-law shape to the 
combined OSSE/EGRET data, they obtained a photon index
of 1.87$\pm$0.02.
We have added the simultaneous COMPTEL data to this spectrum.
Fitting the combined simultaneous OSSE/COMPTEL/EGRET
 spectrum with a simple power-law shape results
in a reduced $\chi^{2}$ of 1.31 and photon spectral
index of 1.84$\pm$0.02 (Table~3). 
If one fits a broken power law model using the formula described 
in Eq.~(2) an improved reduced $\chi^{2}$ of 0.67 (Table~3) is derived.
However, the break parameters cannot be constrained significantly.
The measurements are too uncertain to constrain the spectral shape.

\subsection{Multiwavelength observations}

During the CGRO ToO \gray\ observation, measurements at lower energies
were conducted as well.  
IUE observed the source on July 1, 5,
6, 8 and found time variability of its flux (Bonnell et al., 1995).
Optical observations indicated that the quasar was very active 
in June and July. 
A R-band light curve for the 1995 flaring period is given in Fig.~8.
The source is 3 - 5 magnitudes brighter than previously observed 
(e.g. Torres \& Wroblewski, 1984; Saikia et al., 1987) and  
shows flux variability larger than a factor of 2 within 1 day 
(TJD 9902-9903). 
Due to the sparse sampling in the optical, 
flux correlations between optical and $\gamma$-ray energies cannot be identified.  

%%%%% Fig 8 %%%%%%%%%%%%%%%%%%%%%%%%%%%%%%%%%%%%%%%%%%%%
\begin{figure}[tb]
\centering
\includegraphics[width=8.0cm]{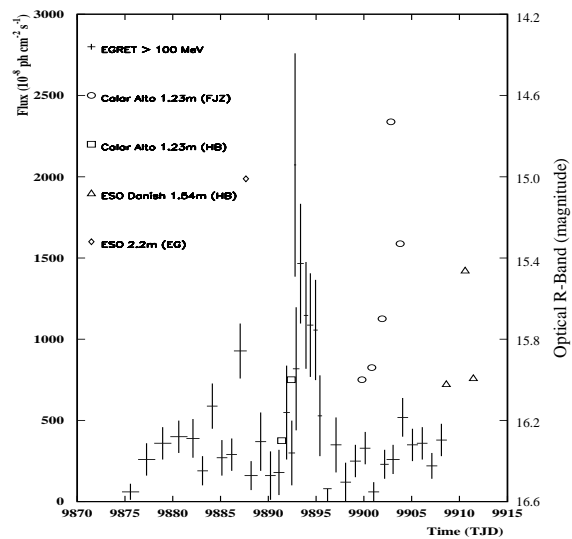}
\caption{Time history of PKS 1622-297 in \grays\ at energies above 100 MeV compared to the optical R band (symbols) during its 5-week flare state in 1995.
The error bars are 1 \sig.
\label{fig:broad_band_light}}
\end{figure} 
%%%%% Fig 8 %%%%%%%%%%%%%%%%%%%%%%%%%%%%%%%%%%%%%%%%%%%%

%%%%% Fig 9 %%%%%%%%%%%%%%%%%%%%%%%%%%%%%%%%%%%%%%%%%%%%
\begin{figure}[tb]
\centering
\includegraphics[width=8.0cm]{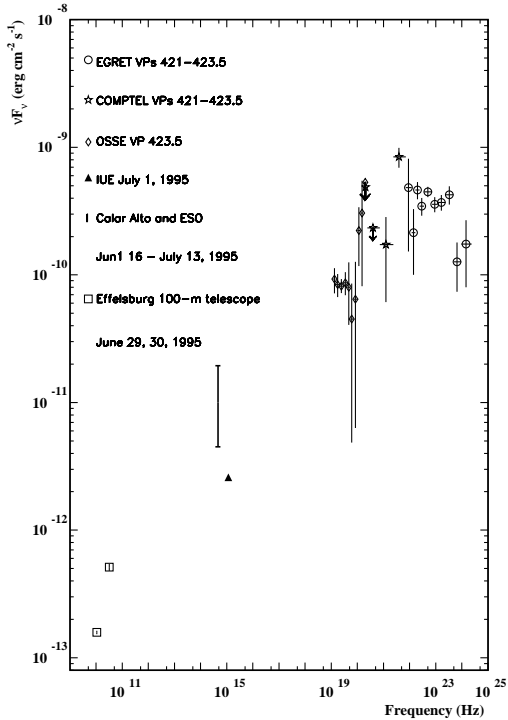}
\caption{Multifrequency spectrum of the flaring quasar PKS 1622-297 from radio to high energy \grays\ during its 5-week flare in 1995.
See details in the text for the individual observations in
the different frequency intervals.  The error bars are 1 \sig\ and the upper limits are 2 \sig. The optical point gives the range between the lowest and the highest R-band fluxes (see Fig.8) during the flaring period of PKS~1622-297 in 1995. 
\label{fig:broad_band_spec}}
\end{figure} 
%%%%% Fig 9 %%%%%%%%%%%%%%%%%%%%%%%%%%%%%%%%%%%%%%%%%%%%

The Effelsberg 100-m radio telescope conducted 4 observations
at the frequency of 10.45 GHz and 1 observation at 32 GHz 
between June 22 and July 20, 1995. No significant flux 
increase compared to previous measurements was found at the
two frequencies (Reich et al., 1998). 
Using these (quasi) simultaneous multiwavelength observations,
we generated the multiwavelength spectrum of PKS~1622-297 during 
its \gray\ high state (Fig.~\ref{fig:broad_band_spec}).
The spectrum shows that the \grays\ are dominating the bolometric 
luminosity by far, and that the luminosity peak is at MeV energies. 
This is similar to other such \gray\ loud flat-spectrum radio quasars
like PKS~0528+134 (e.g. Collmar et al., 1997a) and 3C~279 (e.g. Hartman et al., 2001; Collmar et al., 2001).

\section{Discussion}
%######################################################################

During the first 6 years of the CGRO mission, the radio-loud quasar
PKS 1622-297 was observed several times by the CGRO instruments EGRET
and COMPTEL. \gray\ emission at energies above 100~MeV was detected
by the EGRET experiment in June and July 1995, and also simultaneously 
by COMPTEL at lower 
\gray\ energies; in particular in the 10-30~MeV band. 
An energy flux (10~MeV - 10~GeV) of
2.67$\times$10$^{-9}$ erg cm$^{-2}$ s$^{-1}$ is obtained from the
combined COMPTEL/EGRET spectrum during that period.
For a Friedmann universe with q$_0$=1/2 and 
H$_0$=75~km~s$^{-1}$ Mpc$^{-1}$, 
we derive an isotropic luminosity of 4.47$\times$10$^{48}$ erg s$^{-1}$. The corresponding minimum mass of the putative central black hole for 
Eddington-limited accretion is about 3.55$\times$10$^{10}$ M$_{\odot}$
in the Thompson regime, and about 3.4$\times$10$^8$ M$_{\odot}$ 
when the Klein-Nishina cross section for Compton scattering of
the high energy \gray\ photons is taken into account. 
The limit on the Schwarzschild radius is then given as R$\rm_s$ $>$ 1.1$\times$10$^{14}$ cm, which is a factor of $\sim$ 2 lower than obtained by Mattox  et al. (1997).

Many COMPTEL results suggest that the MeV band is a spectral transition
region for the \gray\ emission of blazars
 (e.g. Hermsen et al., 1993; Bloemen et al., 1995; Collmar et al., 1997a) .
This becomes obvious by multifrequency observations of these sources
which show that their luminosities often peak at MeV energies, 
and -- if no MeV observations are available -- 
by the fact that the soft \gray\ spectra measured by EGRET 
have to match the hard X-ray spectra. 
EGRET revealed that blazars are always variable at \gray\ energies 
(e.g. Mukherjee et al., 1997) . The observational facts of
short-term variability ($\sim$ days)
and simultaneously large luminosities imply a very compact
emission region. 
In order to avoid the problem of \ga-\ga\ absorption in such regions of
extreme photon densities, 
emission of the \grays\ from a relativistic jet 
is generally assumed  (e.g. Blandford \& Rees, 1978).

The current scenario for modeling such blazars involves
 plasma blobs or shocked regions  
filled with charged particles confined within a relativistic jet. 
The \gray\ photons are generated by inverse-Compton (IC) scattering 
of low-energy photons off a population of relativistic electrons
 and positrons.
The primary accelerated particles are still unknown: they can be 
either leptons of a pair-plasma (e.g. Maraschi et al., 1992; Dermer \& Schlickeiser, 1993)
or protons (e.g. Mannheim \& Biermann, 1992). 
For the leptonic models, the soft target photons can
be either self-generated synchrotron photons in the 
so-called synchrotron self-Comptonization (SSC) process
(e.g. Maraschi et al., 1992; Bloom \& Marscher, 1996) 
or these photons can originate external to the jet -- i.e. from 
an accretion disk -- in the so-called 
external Comptonization (EC) process. 
The EC models can be subdivided according to the origin
of the soft photons, into whether the soft photons travel directly
from the accretion disk into the jet
(e.g. Dermer et al., 1992; Dermer \& Schlickeiser, 1993), or whether the soft 
accretion disk photons are first 
scattered by the broad-line region, and hence have an isotropic
distribution (e.g. Sikora et al., 1994). 
Because all mentioned processes are physically valid, 
it is reasonable to consider that they all contribute, 
however with different weights (e.g. B\"ottcher \& Dermer, 1998).
For example, B\"ottcher \& Collmar (1998) have first applied such a modeling to 
explain the spectral variability of the \gray\ blazar PKS 0528+134.
They argued that the SSC component dominates during \gray\ low states
while the EC emission dominates in the flaring periods.
Recently, it was applied to model the multifrequency emission 
of the blazar 3C~279 by Hartman et al. (2001).

PKS~1622-297 fits into the above mentioned scenario. 
During the 5-week flaring state, it showed a variable flux 
at EGRET energies. On top of that EGRET detected a short 
huge flare, showing a rapid flux increase by a factor of 4 in 
less than 7 hours and a corresponding isotropic 
\gray\ luminosity of 2.9$\times$10$^{49}$ erg s$^{-1}$
(Mattox et al., 1997). Both facts strongly argue for Doppler boosted
emission. Mattox et al. (1997), assuming a co-spatial production
of X-rays and \grays, derived an optical
depth for \ga-\ga\ absorption of $\tau_{\gamma\gamma}$ = 330
at 1 GeV. 
Subsequently, the Doppler factor for the bulk motion in the jet then
has to be larger than 8.1.
COMPTEL finds enhanced MeV emission during that 
period, however with the trend of a time delay (see Sect. 4.2). 
The MeV photons may come later than the higher energy photons
due to cooling effects in the relativistic plasma.
The relativistic electrons can loose their energy via
synchrotron or inverse Compton radiation with a
loss rate of $\dot\gamma$ $\propto$ $\gamma^{2}$,
where \ga\ is the Lorentz factor of a given electron.
The characteristic electron life time is
$\Delta\tau$ $\sim$ $\gamma/\gamma^{2}$ $\propto$ $\gamma^{-1}$ and,
in terms of energy E (E $\propto$ $\gamma^{2}$) for
the produced \gray\ photon, is $\Delta\tau$ $\propto$ E$^{-1/2}$
(e.g. Kazanas et al., 1998).
This illustrates that the life time of relativistic electrons
decreases through a cooling process of
 producing higher energies \gray\ photons, while the electron 
population could stay longer by producing lower
 energies \gray\ photons.  Therefore, a time delay of the 
low-energy \gray\ photons seems possible from such
a simple scenario of energy-dependent cooling times 
for a relativistic electron jet.
Numerical calculations show, however, that the case
 of time lags between different energy bands is more complicated.  
Dermer et al. (1999) studied the behavior of blazar flares on 
their time-dependent lags by assuming that a nonthermal plasma is
injected over a finite time interval in the comoving frame and
cools by synchrotron and SSC radiation. 
They argued that the SSC radiation will dominate the 
cooling process if the total energy injected 
in the nonthermal electrons is very high. 
Their numerical results show little time lags between 
the MeV and GeV light curves in the case of
a SSC-dominated cooling process.
Sikora et al. (2001) assumed that the EC process dominates the high
energy \gray\ emission during a flare, and consequently took 
additionally the EC process as cooling mechanism
into account. They considered the \gray\ flare event in two parts:
1) radiative flares caused by a radiative cooling process
like EC emission, 
and 2) adiabatic X-ray and soft \gray\ flares generated 
via adiabatic energy losses of a low-energy electron/positron
population in the jet.
Since adiabatic flares decay much slower than the radiative ones,
the MeV light curve should lag the GeV one. Time lags of several
days should occur between the two energy bands according to 
their numerical simulations.
Our study on the pre-flare, on-flare and post-flare parts of VP 423
provides, for the first time some evidence
 -- though with marginal statistical significance --
for a possible time delay of a few days between the MeV and
$>$100~MeV photons. 
The delay time is consistent with the expected cooling times
in Sikora  et al. (2001).

We observe a 'hard' MeV spectrum of PKS~1622-297 during the flare state,
which indicates that the COMPTEL band is still on the rising part of 
the IC emission. The combined COMPTEL/EGRET spectrum reveals a 
spectral break indicating that for the flaring state 
the IC emission peaks on average around 16~MeV. 
The spectral break is larger than 0.5 in index
which cannot be explained by single-component emission models like SSC
and one-component, time-integrated EC models. 
If compared to the \gray\ quiescent state, a flux increase could be proven
only for the high-energy part of the COMPTEL band. 
This might indicate that the \gray\ flare is mainly a high-energy 
\gray\ phenomenon.  
This \gray\ behaviour resembles the one of PKS~0528+134
(Collmar et al., 1997a), which during 
\gray\ flaring episodes showed a spectral break at MeV energies and 
a power-law shape during \gray\ quiescent states.
This was interpreted as evidence for an additional spectral component
showing up during \gray\ flaring periods. 
Subsequently the \gray\ spectrum of PKS~0528+134 was
modeled by a multi-component scenario consisting of a broad SSC and a narrow 
EC components (B\"ottcher \& Collmar, 1998; Mukherjee et al., 1999).
Very recently such a multi-component model was successfully applied to 
multi-epoch multi-frequency data of the \gray\ blazar 3C~279 
(Hartman et al., 2001), which is a similar source as PKS~1622-297.    
Due to the similarity of the \gray\ behaviour of PKS~1622-297 in the 
flaring state, we shall interpret our measurements also in such a 
multi-component scenario: a combination of SSC and EC processes. 
The measured \gray\ spectrum from 50~keV to 10 GeV for VP 423.5, 
the only one in which OSSE contributed,  provides tentative  
support for such a multi-component scenario. The extrapolation of the 
best-fit OSSE slope significantly underestimates the MeV fluxes, indicating 
such an additional component at MeV energies.  A spectral upturn 
somewhere between  $\sim$~0.5 and 10~MeV is suggested by Fig.~7. 
This agrees well with the model fits on PKS~0528+134 and 3C~279,
which provide a transition from an SSC dominated spectral part to an EC 
dominated one at roughly 0.5~MeV (Mukherjee et al., 1999; Hartman et al., 2001).
However, this indication is not statistically forcing, because
the OSSE slope ($\alpha$: 2.0$\pm$0.2, Mattox et al., 1997) is consistent 
within error bars  with the power-law fit to all data 
($\alpha$: 1.84$\pm$0.02; Fig.~7; Tab.~3). 
   
The SSC and the EC emission processes have different
dependencies on the Doppler 
factor D = ($\Gamma$(1-$\beta$cos$\theta$))$^{-1}$ of the jet, 
where $\Gamma$ = (1-$\beta$)$^{-1/2}$ is the bulk Lorentz factor of 
the relativistic jet plasma, $\beta$ the bulk velocity in units 
of the speed of light, $\theta$ the angle between the direction 
of motion of jet and the line of sight.
According to Dermer et al. (1997), the EC component has a stronger 
dependence ($F_{EC}(\epsilon) \propto D^{3+s}$) on the Doppler factor
than the SSC component ($F_{SSC}(\epsilon) \propto D^{(5+s)/2}$)
and dominates at higher (e.g. MeV) energies (e.g. B\"ottcher \& Collmar, 1998).
The parameter
$s$ is the spectral index of the initial particle distribution function.
For example, if we take $s$ $\sim$ 3 as is generally assumed
 by B\"ottcher \& Collmar (1998) in their modeling of the \gray\ 
emission of PKS 0528+134, then
the dependence of the two components on Doppler factor differs
by a factor of D$^2$. By taking D $>$ 8.1 as was estimated 
by Mattox et al. (1997) for the short huge flare, the difference could
 be $>$ 66. 
Therefore large flux changes can be more efficiently promoted 
by an EC component, which - in the assumed SSC and EC multi-component 
scenario - makes the huge \gray\ flare  at energies $>$ 10 MeV 
during VP~423 likely to be an EC emission event. This interpretation is
supported by the fact, that COMPTEL detects the flare only above 10~MeV,
 which indicates a sharply rising MeV spectrum.

\section{Summary}
%######################################################################

A complete study of PKS 1622-297 using 6 years of COMPTEL
 data is reported in this paper.
 PKS 1622-297 is detected by COMPTEL in its 10-30~MeV 
band as an MeV emitter at a significance level of $~$5.9 \sig\
during a 5-week \gray\ flaring period in 1995.
No evidence for the source is found in the rest of the observations.
Comparing the average flux of the flaring state to the one of the quiescent 
state provides a flux increase by a factor of at least 4
at the upper COMPTEL energies.
During the \gray\ high state, the 5-week average MeV spectrum 
shows a hard shape and a spectral break is indicated by
comparing the MeV spectrum to the EGRET one at energies above 30 MeV.
 The spectral break in the simultaneous COMPTEL/EGRET spectra
suggests that the peak of the IC emission is 
around 16~MeV. In this respect PKS~1622-297 is similar to 
PKS~0528+134 and the same emission scenario might be at work.
A model with two components, SSC and EC, can in principle explain 
the observed \gray\ behaviour.

On top of the 5-week \gray\ activity, EGRET observed a 
huge short ($>$1 day) flare. COMPTEL also detects this event in its
uppermost (10-30 MeV) band. The 10-30 MeV light curve shows that
the MeV emission is still high, when the EGRET flux above 100 MeV
has already ceased, suggesting a time delay between the 2 bands, 
which however is at a marginal significance level.
Modelling \gray\ flare events by assuming that the 
EC process dominates can provide time lags of several days 
(Sikora et al., 2001) in agreement with the indications found in our analysis of PKS~1622-297.

\begin{acknowledgements}
This research was supported by the German government through DLR grant
 50 QV 9096 8, by NASA under contract NAS5-26645, and by the Netherlands
Organization for Scientific Research NWO. The authors would like to thank the anonymous referee for his/her helpful comments.

\end{acknowledgements}


\begin{thebibliography}{}
%######################################################################
\bibitem[Blandford \&  Rees(1978)]{Blandford78}
Blandford, R. D., \& Rees, M. J. 1978, in Pittsburgh Conference on BL Lac Objects, A.~M.~Wolfe, Univ. Pittsburgh Press, 328

\bibitem[Bloemen et al.(1995)]{Bloemen95}
Bloemen, H., Bennett, K., Blom, J. J., et al. 1995, A\&A, 293, L1

\bibitem[Bloemen et al.(1994)]{Bloemen94}
Bloemen, H., Hermsen, W., Swanenburg, B. N., et al. 1994, ApJS, 92, 419

\bibitem[Blom et al.(1995a)]{Blom95a}
Blom, J. J., Bloemen, H., Bennett, K., et al. 1995a, A\&A, 295, 330

\bibitem[Blom et al.(1995b)]{Blom95b}
Blom, J. J., Bloemen, H., Bennett, K., et al. 1995b, A\&A, 298, L33

\bibitem[Bloom \& Marscher(1996)]{Bloom96}
Bloom, S. D., \& Marscher, A. P. 1996, ApJ, 461, 657

\bibitem[de Boer et al.(1992)] {Boer92}
de Boer, H., Bennett, K., Bloemen, H., et al. 1992, in Data Analysis in Astronomy IV, eds. V. Di Ges\`{u}, L. Scarsi, R. Buccheri, et al., (New York: plenum Press), 241

\bibitem[Bonnell et al.(1995)] {Bonnell95}
Bonnell,  J. T., Shrader, C. R., Hartman, R. C., et al. 1995, IAU circular 6186

\bibitem[B\"ottcher \& Dermer(1998)]{Bottcher98a}
B\"{o}ttcher, M., \& Dermer, C. D. 1998, ApJ, 501, L51

\bibitem[B\"ottcher \& Collmar(1998)]{Bottcher98b}
B\"{o}ttcher, M., \& Collmar, W. 1998, A\&A, 329, 57


\bibitem[Collmar et al.(1997a)]{Collmar97a}
Collmar, W., Bennett, K., Bloemen, H., et al. 1997a, A\&A, 328, 33

\bibitem[Collmar et al.(1997b)]{Collmar97b}
Collmar, W., Bennett, K., Bloemen, H., et al. 1997b, in Proceedings of the 
Fourth Compton Symposium, eds. C. D. Dermer, M. S. Strickman, J. D. Kurfess (New York: AIP Conf. Proc. 410), 1341  

\bibitem[Collmar et al.(2000)]{Collmar00}
Collmar, W., Reimer, O., Bennett, K., et al. 2000, A\&A, 354, 513

\bibitem[Collmar et al.(2001)]{Collmar01}
Collmar, W., Sch\"onfelder, V., Zhang, S., et al. 2001, in Proceedings of the 
$\gamma$-Ray Astrophysics 2001, eds. S. Ritz, N. Gehrels, C. R. Shrader (New York: AIP Conf. Proc. 587), 271  


\bibitem[Dermer et al.(1997)]{Dermer97}
Dermer, C. D., Sturner, S. J., \& Schlickeiser, R. 1997, ApJS, 109, 103 

\bibitem[Dermer et al.(1992)]{Dermer92}
Dermer, C. D., Schlickeiser, R., \& Mastichiadis, A. 1992, A\&A, 256, L27

\bibitem[Dermer \& Schlickeiser(1993)]{Dermer93}
Dermer, C. D., \& Schlickeiser, R. 1993, ApJ, 416, 458

\bibitem[Dermer et al.(1999)]{Dermer99}
Dermer, C. D., Li, H., \&  Chiang, J. 1999, Astro. Lett. and Communications, 39, 1


\bibitem[Hartman et al.(1999)] {Hartman99}
Hartman, R. C., Bertsch, D. L., Bloom, S. D., et al. 1999, ApJS, 123, 79

\bibitem[Hartman et al.(2001)] {Hartman01}
Hartman, R. C., B\"ottcher, M., Aldering, G., et al. 2001, ApJ, 553, 683

\bibitem[Hermsen et al.(1993)] {Hermsen93}
Hermsen, W., Aarts, H. J. M., Bennett, K., et al. 1993, A\&AS, 97, 97

\bibitem[Impey \& Tapia(1990)] {Impey90}
Impey, C. D., \& Tapia, S. 1990, ApJ, 354, 124

\bibitem[Kazanas et al.(1998)] {Kazanas98}
Kazanas, D., Titarchuk, L. G., \& Hua, X. M. 1998, ApJ, 493, 708


\bibitem[K\"uhr et al.(1981)] {Kuhr81}
K\"{u}hr, H., Witzel, A., Pauliny-Toth, I. I. K., \& Nauber, A. 1981, A\&AS, 45, 367

\bibitem[Kurfess et al.(1995)] {Kurfess95}
Kurfess, J. D., Grove, J. E., McNaron-Brown, K., et al. 1995, IAU circular 6185

\bibitem[Lampton et al.(1976)] {Lampton76}
Lampton, M., Margon, B., \& Bowyer, S. 1976, ApJ, 208, 177

\bibitem[Mannheim \& Biermann(1992)]{Mannheim92}
Mannheim, K., \& Biermann, P. L. 1992, A\&A, 253, L21

\bibitem[Maraschi et al.(1992)]{Maraschi92}
Maraschi, L., Ghisellini, G., \& Celotti, A. 1992, ApJ, 397, L5

\bibitem[Mattox et al.(1997)] {Mattox97}
Mattox, J. R., Wagner, S. J., Malkan, M., et al. 1997, ApJ, 476, 692

\bibitem[Mukherjee et al.(1997)] {Mukherjee97}
Mukherjee, R., Bertsch, D. L., Bloom, S. D., et al. 1997, ApJ, 490, 116

\bibitem[Mukherjee et al.(1999)] {Mukherjee99}
Mukherjee, R., B\"ottcher, M., Hartman, R. C., et al. 1999, ApJ, 527, 132


\bibitem[Reich et al.(1998)] {Reich98}
Reich, W., Reich, P., Pohl, M., Kothes, R., \& Schlickeiser, R. 1998, A\&AS, 131, 11

\bibitem[Saikia et al.(1987)] {Saikia87}
Saikia, D. J., Ashok, K. S., \& Cornwall, T. T. 1987, MNRAS, 224, 379

\bibitem[Sch\"onfelder et al.(1993)]{Schonfelder93}
Sch\"onfelder, V., Aarts, H., Bennett, K., et al. 1993, ApJS, 86, 657 

\bibitem[Sikora et al.(1994)] {Sikora94}
Sikora, M., Begelman, M. C., \& Rees, M. J.  1994, ApJ, 421, 153   

\bibitem[Sikora et al.(2001)] {Sikora01}
Sikora, M., Blazejowski, M., Madejski, G., \& Moderski R.  2001, in Proceedings of the 4th Integral Workshop, eds. A. Gimenez, V. Reglero, C. Winkler (ESA SP-459), 259  


\bibitem[Steppe et al.(1993)] {Steppe93}
Steppe, H., Paubert, G., Sievers, A., et al. 1993, A\&AS, 102, 611


\bibitem[Torres \& Wroblewski(1984)] {Torres84}
Torres, C., \& Wroblewski, H. 1984, A\&A, 141,271; and erratum, A\&A, 160, 406

\bibitem[Williams et al.(1995)]{Williams95}
Williams, O. R., Bennett, K., Bloemen, H., et al. 1995, A\&A, 298, 33

\bibitem[Wright \& Otrupcek(1990)] {Wright90}
Wright, A., \& Otrupcek, R. 1990, PKSCTA90 catalog


\end{thebibliography}
\end{document}